\documentclass[aps,prb,preprint,groupedaddress]{revtex4-1}
\usepackage[T1]{fontenc}
\usepackage[latin9]{inputenc}
\usepackage{amssymb}
\usepackage{graphicx}
\usepackage{bm}

\def\be{\begin{equation}}
\def\ee{\end{equation}}
\def\bea{\begin{eqnarray}}
\def\eea{\end{eqnarray}}

\begin{document}

\title{The crater function approach to ion-induced nanoscale pattern formation: Craters for flat surfaces are insufficient}

\author{Matt P. Harrison and R. Mark Bradley}
\affiliation{Department of Physics, Colorado State University, Fort
Collins, CO 80523, USA}

\date{\today}

\begin{abstract}

In the crater function approach to the erosion of a solid surface by a broad ion beam, the average crater produced by the impact of an ion is used to
compute the constant coefficients in the continuum equation of motion for the surface.
We extend the crater function formalism so that it includes the dependence of the crater 
on the curvature of the surface at the point of impact.  We then demonstrate that our formalism yields the correct coefficients for the Sigmund model of ion sputtering if terms up to second order in the spatial derivatives are retained.  In contrast, if the curvature dependence of the crater is neglected, the coefficients can deviate substantially from their exact values.  Our results 
show that accurately estimating the coefficients using craters obtained 
from molecular dynamics simulations will require significantly more computational power than was previously thought.

\end{abstract}

\date{\today}

\maketitle

\section{Introduction}
\label{sec:introduction}

Bombarding a solid surface with a broad ion beam can lead to the spontaneous formation of nanoscale patterns on the surface.\cite{Munoz-Garcia09a}  These patterns include 
periodic height modulations or \lq\lq ripples'' as well as nanodots arranged in hexagonal arrays of surprising regularity.\cite{Facsko99,Frost00,Wei09,Fritzsche12,Bischoff11a,Bischoff11b}
This has spurred widespread interest in the development of ion sputtering as a means of nanofabrication.  Since broad beam ion bombardment is relatively easy to implement, the potential for cost-effective mass production of nanostructures is quite high.

Much of the theoretical work done in analyzing these patterns has been based on the continuum Bradley-Harper (BH) theory,$\cite{Bradley88}$ which itself is based on the Sigmund model of ion sputtering.\cite{Sigmund73}  BH showed that for the Sigmund model the sputter yield at a point on the surface does not just depend on the local angle of incidence --- it depends on the surface curvature as well.  Because high points on the surface are eroded more slowly than the low points, the curvature dependence of the sputter yield leads to an instability of the solid surface. The BH theory has been extended to include nonlinear effects\cite{Cuerno95,Makeev02,Castro05,Munoz-Garcia08} and so that it applies to binary materials.\cite{Shenoy07}

Since the work of Carter and Vishnyakov (CV) in 1996,\cite{Carter96} it has become increasingly clear that ion-induced mass redistribution can play an important role in the pattern formation.\cite{Moseler05,Davidovitch07,Kalyanasundaram08,Kalyanasundaram09,Madi11,Norris11,Castro12a,
Norris12a,Norris12b,Castro12b,Bobes12,Hofsass13a,Yang13,Moller14} In this process, momentum is transferred from the incident ions to atoms near the surface of the solid.  These atoms are not ejected from the solid surface as they would be in sputtering.  
Instead, they are displaced within the solid.

The theories of BH and of CV are based on simple models of sputtering and mass redistribution.  It has been unclear just how good these models are and in what circumstances they can be reasonably applied.  Moreover, the predictions of the BH and CV theories depend on a number of phenomenological parameters but give no means of computing their values.

Recently, there has been considerable interest in incorporating the results of molecular dynamics (MD) simulations into a continuum theory of ion-induced surface dynamics.  The so-called crater function formalism (CFF) utilizes the average result of many ion impacts at a single point to generate a Green's function, which is then used to determine the response of a surface to bombardment with a broad ion beam.$\cite{Norris09,Norris11}$  This approach has the advantage that it takes into account both sputtering and ion-induced mass redistribution and does not rely on simple models of these phenomena.  The formalism yields estimates of the constant coefficients that appear in the continuum equation of motion based on input from MD simulations.

In the first application of this method to a specific physical problem, Norris {\it et al.}~carried out MD simulations of the bombardment of a silicon surface with 100 and 250 eV Ar$^+$ ions and then used their CFF to obtain estimates of some of the coefficients in the equation of motion.\cite{Norris11}  Based on these results, they concluded that mass redistribution is predominant and that the curvature dependence of the sputter yield is \lq\lq essentially irrelevant.''  They then went further and declared that this \lq\lq conclusion overturns the erosion based paradigm that has dominated the field for two decades,'' even though their results were restricted to ion bombardment of a single material with low energy ions of a particular species.

The Green's function, which is usually referred to as the \lq\lq crater function,'' 
depends on the complete shape of the surface surrounding the impact point.\cite{Norris09}  However,
because it is not possible to find the crater function for an arbitrarily shaped surface using MD, 
the shape dependence of the crater was simply neglected in Norris {\it et al.}'s study of the erosion of Si with an Ar$^+$ beam.\cite{Norris11}
In particular, the crater function for a \emph{flat} surface was used to estimate the coefficients in the equation of motion (EOM),
even though the accuracy of such a procedure is questionable.  The dependence of the crater on the shape of the surface has also been neglected in more recent applications of the CFF.\cite{Hossain11,NorrisXX}

In this paper, we extend the CFF so that it includes the dependence of the crater function on the curvature of the surface at the point of impact.  We give
explicit expressions for the coefficients in the equation of motion which reduce to the expressions given by Norris {\it et al.}\cite{Norris11}~if the curvature dependence of
the crater function is neglected.  We then demonstrate that our extended CFF yields the exact BH coefficients for the Sigmund model.  
In contrast, the BH coefficients are not recovered if the curvature dependence of the crater function is neglected.  This uncontrolled approximation instead results in coefficients that are off by a factor of two for normal-incidence bombardment.  Norris {\it et al.}'s estimated coefficients for bombardment of Si with an Ar$^+$ beam led to their
overarching claim that mass redistribution is always much more important than the curvature dependence of the sputter yield, 
but our results cast doubt on the reliability of these estimates.

This paper is organized as follows.  We introduce the crater function and its arguments in Section II.  In Section III, we use the crater function to determine the coefficients in the EOM for the special case in which the surface height does not vary in the direction transverse to the plane of the beam.  In Section IV, we develop the geometric preliminaries required to extend our theory to fully three-dimensional surfaces.  Section V generalizes the results of Section III to the case in which the surface height varies in both the transverse and longitudinal directions.  Section VI contains an explicit demonstration that our extended CFF is in accord with the BH theory in the case of the Sigmund crater.  In Section VII, we compare our theory to the CFF of Norris {\it et al.}\cite{Norris11}~and demonstrate that for the Sigmund crater the latter produces coefficients that can differ significantly from their exact values.  Additionally,
we discuss the implications of our work, and place its results in context.  Our findings are summarized in Section VIII. 

\section{The Crater Function}
\label{sec:eoms}

Consider the bombardment of a solid elemental material with a broad ion beam.  We will assume that the material is amorphous, or, if it is crystalline, that a layer at the surface of the solid is rendered amorphous by the ion bombardment.  The sample surface will be taken to be nominally flat before the irradiation begins.

We define the $\bm{\hat{z}}$ direction to be the global vertical, normal to the macroscopic surface.  $\bm{\hat{x}}$ is taken to be the direction of the projection of the incident ion beam onto the macroscopic surface, and $\bm{\hat{y}}$ is taken to be normal to the $x-z$ plane.  The incident ion flux is $\bm{J}=J(\bm{\hat{x}}\sin{\theta}-\bm{\hat{z}}\cos{\theta})$, where the angle of incidence $\theta$ is the angle between the global vertical and the incident beam, as shown in Fig.~1.  An arbitrary point on the surface \textbf{P} is given by $\bm{r}=x\bm{\hat{x}}+y\bm{\hat{y}}+h(x,y)\bm{\hat{z}}$, where $h(x,y)$ is the height of the point above the $x-y$ plane.  (For convenience, we will suppress the time dependence of $h$ unless it is necessary to explicitly display it.)  

Our goal is to evaluate ${\partial h}/{\partial t}$ at an arbitary point $\textbf{O}$ on the solid surface at an arbitrary time $t>0$.  To that end, we will place the global origin at the position
of $\textbf{O}$ at time $t$, as shown in Fig.~1.  The global origin will be taken to be stationary, and it so will remain fixed as the surface point $\textbf{O}$ moves either up or down.

\begin{figure}[h]
\begin{centering}
\includegraphics[width=3.0in]{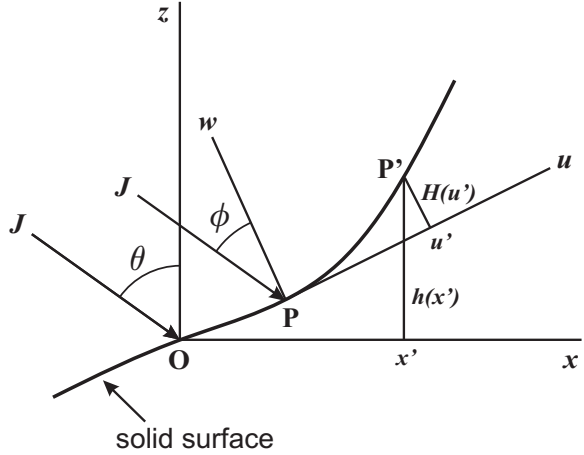}
\caption{The solid surface at time $t$.  The points \textbf{O}, \textbf{P} and \textbf{P'} lie on the surface.  The global frame of reference has its 
origin at \textbf{O} and has axes $x$, $y$ and $z$, while the local frame of reference has its origin at \textbf{P} and has axes $u$, $v$ and $w$. $\bm{J}$ is the incident ion flux.
$\theta$ and $\phi$ are the global and local angles of incidence, respectively.  The height of the point \textbf{P'} is $h(x')$ in the global frame but is $H(u')$ in the local frame.  For simplicity, the figure has 
been drawn for the special case in which $h(x,y)$ is independent of $y$.}
\par\end{centering}
\end{figure}

The collision cascade that an impinging ion produces in the solid has a characteristic lateral length scale that we will denote by $l$. We will assume that a smoothing mechanism ensures that the surface height varies only a
little over this length scale; in practice, the smoothing mechanism could be thermally activated surface diffusion (as in the BH theory)
or ion-induced viscous flow.\cite{Umbach01}
It is important to note that the equation of motion we will derive will \emph{not} include the effects of the smoothing mechanism, since we will include
only terms up to second order in the wave number $k$ and the smoothing mechanism produces terms of order $k^4$.

Our first step in finding the surface velocity at $\textbf{O}$ will be to determine the contribution to it coming from ions striking the surface an arbitrary surface point \textbf{P}.  
In fact, we may restrict our attention
to points \textbf{P} that have a distance to \textbf{O} that is on the order of a few times $l$ or less because ions arriving at more remote points make a negligible contribution to the value of ${\partial h}/{\partial t}$ for $x=y=0$.  The height $h$ is small for these points \textbf{P}.   We will accordingly work to first order in $h$ and its spatial derivatives throughout the remainder of the paper.

In addition to the global coordinates $x$, $y$ and $z$, it is convenient to introduce a set of local coordinates whose origin is the point \textbf{P}.  Following Norris, Brenner and Aziz,\cite{Norris09} we define the vector $\bm{\hat{n}}$ to be the local surface normal at \textbf{P} and $\bm{\hat{t}}_u$ to be the local downbeam direction projected onto the surface.  Explicitly,
\be
\bm{\hat{n}}=\frac{\bm{\hat{z}}-\bm{\nabla} {h}}{\sqrt{1+(\nabla {h})^2}}
\ee
and
\be
\bm{\hat{t}}_u=\frac{-\bm{J}+(\bm{J}\cdot\bm{\hat{n}})\bm{\hat{n}}}{|-\bm{J}+(\bm{J}\cdot\bm{\hat{n}})\bm{\hat{n}}|}.
\ee
$\bm{\hat{t}}_v$ is defined to be the cross product of $\bm{\hat{n}}$ and $\bm{\hat{t}}_u$.  The unit vectors $\bm{\hat{n}}$, $\bm{\hat{t}}_u$ and $\bm{\hat{t}}_v$ form an orthonormal basis and $\bm{\hat{t}}_u$ and $\bm{\hat{t}}_v$ are tangent to the surface at \textbf{P}.  The local angle of ion incidence, which will be denoted by $\phi$, is given by $J\cos{\phi}=-\bm{J}\cdot\bm{\hat{n}}$. 
To first order in the spatial derivatives of the surface height, 
\be
\phi(x,y)=\theta-h_{x}(x,y), \label{phi}
\ee
where the subscript denotes a partial derivative with respect to $x$.  Finally, we define $u$, $v$, and $w$ to be the coordinates along the directions $\bm{\hat{t}}_u$, $\bm{\hat{t}}_v$ and $\bm{\hat{n}}$, respectively.

For surface points that have a distance to \textbf{O} that is on the order of $l$, we may approximate $h$ by discarding terms of third order and higher terms from its Taylor series: We set $x_1=x$, $x_2=y$, and
\be
h(x,y)= S_1x+S_2 y+\frac{1}{2}K_{11}x^2+K_{12}xy+ \frac{1}{2}K_{22}y^2\label{hdef},
\ee
where
\be
S_i\equiv\frac{\partial h}{\partial x_i}(0,0)
\ee
and
\be
K_{ij}\equiv\frac{\partial^2h}{\partial x_i \partial x_j}(0,0) 
\ee
for $i,j=1,2.$  While an arbitrary number of terms in the expansion (\ref{hdef}) could in principle be retained, we will only keep terms up to quadratic order in $x$ and $y$ because the length scale of the height variation is assumed to be much larger than $l$.  Note that the quantities $A_i$ and $K_{ij}$ are both of first order in $h$.  This will be exploited later in our analysis.

We may also parameterize the surface in terms of the local coordinates $u$, $v$, and $w$.  Close to \textbf{P}, the height of the solid surface above the $u-v$ plane is given by 
\be
H(u,v)=\frac{1}{2}E_{11}u^2+E_{12}uv+\frac{1}{2}E_{22}v^2,\label{Hexp}
\ee
to second order in $u$ and $v$.  Here
\be
E_{ij}\equiv\frac{\partial^2H}{\partial u_i \partial u_j}(0,0), \label{EZE}
\ee
where $u_1\equiv u$, $u_2\equiv v$ and $i,j=1,2$.  Terms that are linear in $u$ and $v$ do not appear on the right-hand side of Eq.~(\ref{EZE}) because the $u$ and $v$ axes are
tangent to the solid surface at the point \textbf{P}.  The expansion~(\ref{Hexp}) gives a good approximation to the value of $H$ for \textbf{O} because the distance between \textbf{O} and \textbf{P}
is of order $l$.

We now introduce the crater function
\be
F=F(u,v,\phi,E_{11},E_{12},E_{22}),\label{crater}
\ee
which is defined to be minus the average change in the local surface height $H$ above the point $(u,v)$ in the $u-v$ plane as a result of a single ion impact at $u=v=0$, i.e., the point \textbf{P}.  While two impacts may produce very different craters, by taking the statistical average of a great number of craters, we develop an expected response. The information required to construct $F$ is assumed to be known $\emph{a priori}$ from another theory or from MD simulations.  

The crater function $F(u,v,\phi,E_{11},E_{12},E_{22})$ is defined in the local coordinate system of the point of impact \textbf{P}.  Its first two arguments are the lateral coordinates $u$ and $v$ in that coordinate system.  The third argument of $F$ is the local angle of incidence $\phi$.  Finally, we have included the dependence
of the crater on the local curvatures $E_{11}$, $E_{12}$ and $E_{22}$.  This dependence was neglected by Norris {\it et al.},\cite{Norris11}
but, as we will discuss in Section \ref{sec:discussion}, evidence from experiments\cite{PerkinsonXX} and MD simulations\cite{Nietiadi13} suggests that it can have a significant effect.

Note that while the $E_{ij}$'s refer to second derivatives of $H$ with respect to the local coordinates $u$ and $v$ at the point $\textbf{P}$, it is shown in Section IV that to first order they are equal to the corresponding second derivatives of $h$ with respect to the global coordinates $x$ and $y$ at the point $\textbf{O}$, i.e.,
\be
E_{ij}=K_{ij}
\label{EK}
\ee 
for $i,j=1,2$.  We may therefore rewrite Eq.~($\ref{crater}$) as
\be
F=F(u,v,\phi,K_{11},K_{12},K_{22}).
\ee

\section{The Extended Crater Function Formalism in Two Dimensions}
\label{sec:eoms}

The goal of our analysis is to derive an EOM of the form
\bea
\frac{1}{J}\frac{\partial h}{\partial t}= &&\, C_0(\theta)+C_1(\theta)h_x+C_2(\theta)h_y\nonumber\\&&+C_{11}(\theta)h_{xx}+C_{12}(\theta)h_{xy}+C_{22}(\theta)h_{yy}\label{OGEOM},
\eea
and to write the coefficients $C_0$, $C_1,\ldots , C_{22}$ in terms of the crater function $F$.  The first step in our analysis will be to determine the contribution to the normal velocity of the surface at $\textbf{O}$ due to impacts at the point $\textbf{P}$.  Having found this, we will perform a flux weighted integral over all possible impact points \textbf{P} to determine the overall response.  

To make the analysis as transparent as possible, we will begin by considering the special case in which the surface height $h$ has no dependence on $y$.  
In this case, Eq.~(\ref{OGEOM}) reduces to 
\be
\frac{h_t}{J}=C_0(\theta)+C_1(\theta)h_x +C_{11}(\theta)h_{xx},
\label{OGEOM2}
\ee
where $h_t\equiv {\partial h}/{\partial t}$.  This problem is equivalent to a two-dimensional (2D) problem in which $h$ depends only on $x$ and $t$ and ions are incident in the $x-z$ plane 
with an angle of incidence $\theta$. The effective crater function for this 2D problem is
\be
g(u,\phi,E_{11})\equiv \int_{-\infty}^\infty F(u,v,\phi,E_{11},0,0)dy.
\ee
We will study the equivalent 2D problem for the remainder of this section.

Consider an impact at the point \textbf{P} whose position in the global coordinate system is $\bm{r}=x\bm{\hat{x}}+h(x)\bm{\hat{z}}$.  The lateral position of the global origin \textbf{O} in the local reference frame of the impact point is to first order
\be
u=\bm{\hat{t}}_u(x)\cdot(\bm{0}-\bm{r})=[\bm{\hat{x}}+h_x(x)\bm{\hat{z}}]\cdot[-x\bm{\hat{x}}-h(x)\bm{\hat{z}}]=-x.
\ee
Thus, to first order, we may replace the first argument of the crater function $g(u,\phi,E_{11})$ by $-x$.  Similarly, the height of the origin $\textbf{O}$ relative to the local frame of the impact point \textbf{P} is to first order 
\be
H(u)\equiv\bm{\hat{n}}(x)\cdot(\bm{0}-\bm{r}) = [-h_x(x)\bm{\hat{x}}+\bm{\hat{z}}]\cdot[-x\bm{\hat{x}}-h(x)\bm{\hat{z}}] =xh_x(x)-h(x).
\ee

Recall that the crater function gives the change in surface height in the direction of the local normal $\bm{\hat{n}}$, and so we must project the local normal velocity along the global vertical direction in order to find the velocity of the surface point $\textbf{O}$ along the global vertical direction.  However, because
\be
\bm{\hat{n}}(x)\cdot\bm{\hat{z}}=1
\ee
to first order, this projection has no effect on the linearized EOM we will obtain.

This analysis permits us to write the time derivative of the surface height at \textbf{O} in terms of the crater function $g$ and the ion flux $J$:
\be
h_t(0,t)=-J\int g(-x,\phi,E_{11})\cos{\phi}dx \label{xEOMraw},
\ee
where the factor of $\cos\phi$ comes from projecting the ion flux onto the local normal at the point $\textbf{P}$.  Finally, because only
points \textbf{P} within a distance on the order of $l$ from the origin give a significant contribution to the integral on the right-hand side of
Eq.~(\ref{xEOMraw}), we may replace $E_{11}$ by $K\equiv K_{11}$ in the integral.

We are now in a position to begin analyzing the integrand in Eq.~(\ref{xEOMraw}).  
To do so, we will linearize in the quantities $S\equiv S_1$ and $K$, which, as we noted earlier, are first order in $h$.  This will yield expressions for the coefficients in the EOM (\ref{OGEOM2}).  Making use of $\phi=\theta-h_x=\theta-S-Kx$, we see that
\bea
-J^{-1}h_t(0,t)=&&\int g(-x,\theta,0)\cos{\theta}dx\nonumber\\ &&+S\Bigg[\frac{d}{dS}\int  g(-x,\theta-S,0)\cos{(\theta-S})dx\Bigg]\Bigg|_{S=0}\nonumber\\ &&+K \Bigg[\frac{d}{dK}\int g(-x,\theta-Kx,K)\cos{(\theta-Kx)}dx\Bigg]\Bigg|_{K=0}\label{xEOM}.
\eea

The first term on the right-hand side of Eq.~(\ref{xEOM}) is particularly simple, and gives the steady-state erosion velocity.  Notice that we may perform a change of variable $x\rightarrow -x$ without changing the overall sign of this term, i.e.,
\be
\int g(-x,\theta,0)\cos{\theta}dx=\int g(x,\theta,0)\cos{\theta}dx.
\ee
Therefore, the steady-state erosion velocity for the undisturbed flat surface is 
\be
V_0(\theta) = J\cos{\theta}\int  g(x,\theta,0)dx.\label{steady}
\ee

The second term on the right-hand side of Eq.~(\ref{xEOM}) is somewhat more involved.  Noticing that the only dependence of $g$ upon $S$ comes from the the local angle of incidence $\phi$, it is clear that we may write the second term on the right-hand side of Eq.~(\ref{xEOM}) as
\be
S\Bigg[\frac{d}{dS}\int  g(-x,\theta-S,0)\cos{(\theta-S})dx\Bigg]\Bigg|_{S=0}=-S\frac{\partial}{\partial \theta}\int  g(x,\theta,0)\cos{\theta}dx=-\frac{S}{J}\frac{\partial}{\partial \theta}V_0(\theta).\label{Acoef}
\ee

Finally, we turn to the dependence of $h_t$ on $K$.  The last term on the right-hand side of Eq.~(\ref{xEOM}) becomes
\bea
K \Bigg[\frac{d}{dK} \int &g&(-x,\theta-Kx,K)\cos{(\theta-Kx)}dx\Bigg]\Bigg|_{K=0}\nonumber\\=K\int dx\Big[-x\sin{\theta}&g&(x,\theta,0)+x\cos{\theta}\frac{\partial g}{\partial \theta}(x,\theta,0)+\cos{\theta}\frac{\partial g}{\partial K}(x,\theta,K)\Big|_{K=0}\Big]\label{1DK},\label{Kcoef}
\eea
where we have once again used the change of variable $x\rightarrow-x$.  

Inserting Eqs.~(\ref{steady}), (\ref{Acoef}) and (\ref{Kcoef}) into Eq.~(\ref{xEOM}), we arrive at an EOM of the form (\ref{OGEOM2}).  Defining
\be
M_{K}(\theta)=\int g(x,\theta,K)dx
\ee
and
\be
M^{(n)}_{x}(\theta)=\int g(x,\theta,0) x^{n}dx,
\ee
we obtain
\bea
h_{t}(0,t)=&&-JM^{(0)}_{x}\cos{\theta}+J\frac{\partial}{\partial \theta} (M^{(0)}_{x}\cos{\theta})h_x(0,t)\nonumber\\&&-J\Bigg[\frac{\partial}{\partial \theta}(M^{(1)}_{x}\cos{\theta})+\cos{\theta}\frac{\partial}{\partial K} M_{K}\Big|_{K=0}\Bigg]h_{xx}(0,t).
\eea
Comparing this to Eq.~(\ref{OGEOM2}), we see that
\be
C_0(\theta)=-M^{(0)}_{x}\cos{\theta},\label{C0}
\ee
\be
C_1(\theta)=\frac{\partial}{\partial \theta} (M^{(0)}_{x}\cos{\theta})=-\frac{\partial}{\partial \theta}C_0(\theta),
\label{C1}
\ee
and
\be
C_{11}(\theta)=-\frac{\partial}{\partial \theta}(M^{(1)}_{x}\cos{\theta})-\cos{\theta}\frac{\partial}{\partial K_{11}} M_{K_{11}}\Big|_{K_{11}=0}.
\label{C11}
\ee
The first term on the right-hand side of Eq.~(\ref{C11}) stems from the fact that a nonzero surface curvature gives rise to a local angle of ion incidence
that depends on the point of impact.  The second is a direct result of the curvature dependence of the crater function itself.

\section{Geometric Preliminaries In Three Dimensions}
\label{sec:analysis}

The extension of the analysis of the previous section to three dimensions (3D) is subtle and requires care.  In this section, we delve into the relationship between the local and global coordinate systems before turning to the CFF in 3D.  As discussed in Section II, the local coordinate system is defined using the local surface normal and the projection of the ion beam onto the local tangent plane. 

To first order in $h$, the local unit vectors may be expressed in terms of their global counterparts as follows:
\be
\bm{\hat{t}}_u=\bm{\hat{x}}-(h_y \cot \theta) \bm{\hat{y}}+h_x \bm{\hat{z}}\label{thatu},
\ee
\be
\bm{\hat{t}}_v=(h_y\cot \theta) \bm{\hat{x}}+\bm{\hat{y}}+h_y\bm{\hat{z}}\label{thatv},
\ee
and
\be
\bm{\hat{n}}=-h_x\bm{\hat{x}}-h_y\bm{\hat{y}}+\bm{\hat{z}}\label{nhat}.
\ee
The partial derivatives of $h$ are to be evaluated at the point $(x,y)$ in the $x-y$ plane in these expressions.  The coordinates of the point \textbf{O} in the local coordinate system ($u$, $v$ and $w$) can now be found using Eqs.~(\ref{thatu}) - (\ref{nhat}).  The vector leading from \textbf{P} to \textbf{O}
is $-\bm{r}$.  Recalling that $\bm{r}=x\bm{\hat{x}}+y\bm{\hat{y}}+h(x,y)\bm{\hat{z}}$, we obtain
\be
u=-\bm{r}\cdot\bm{\hat{t}}_u=-x+yh_y\cot \theta\label{uo},
\ee
\be
v=-\bm{r}\cdot\bm{\hat{t}}_v=-y-xh_y\cot \theta\label{vo},
\ee
and
\be
w=-\bm{r}\cdot\bm{\hat{n}}=x h_x + y h_y - h\label{wo}
\ee
to first order.  We may use Eq.~(\ref{hdef}) to eliminate $h$ from Eqs.~(\ref{uo}) - (\ref{wo}) because the surface height varies slowly between 
\textbf{O} and \textbf{P}.  In particular, Eq.~(\ref{wo}) yields 
\be
w =\frac{1}{2}K_{11}x^2+K_{12}xy+\frac{1}{2}K_{22}y^2.
\label{weqn}
\ee

We are now prepared to demonstrate that Eq.~(\ref{EK}) is valid.  Inversion 
of Eqs.~(\ref{uo}) and (\ref{vo}) gives
\be
x=-u-v h_y\cot\theta
\ee
and
\be
y=-v+u h_y\cot\theta.
\ee
Since $H=w$ and the $K_{ij}$'s are first order in $h$, Eq.~(\ref{weqn}) may now be written
\bea
H(u,v)= &&\,\frac{1}{2}K_{11}(u+vh_y\cot\theta )^2+K_{12}(u+v h_y\cot\theta)(v-u h_y\cot\theta)\nonumber\\
&&+\frac{1}{2}K_{22}(v-u h_y \cot\theta)^2\nonumber\\
= &&\,\frac{1}{2}K_{11}u^2+K_{12}uv+\frac{1}{2}K_{22}v^2.\label{Hdef}
\eea
Taking the partial derivatives of $H$ with respect to $u_i$ and $u_j$, we arrive at the desired result, Eq.~(\ref{EK}).

\section{The Extended Crater Function Formalism in Three Dimensions}
\label{sec:discussion}

We will now utilize the results of Section \ref{sec:analysis} to obtain the coefficients of the EOM in three dimensions.
To extend the formalism to the general case in which the surface height depends on $y$ as well as $x$, 
we return to the crater function $F(u,v,\phi,E_{11},E_{12},E_{22})$, the generalization of $g(u,\phi,E_{11})$
to three dimensions.  The EOM is 
\be
h_t=-J\int dx\int dy \cos\phi F(u,v,\phi,E_{11},E_{12},E_{22}).
\ee
Using Eqs.~(\ref{phi}), (\ref{EK}), (\ref{uo}) and (\ref{vo}), we see that this may be written
\be
{h_t}=-{J}\int dx \int dy \cos(\theta-h_x) F(-x+yh_y \cot\theta,-y-xh_y \cot\theta,\theta-h_x,K_{11},K_{12},K_{22}).
\ee
We now expand this to linear order in $h$ and its derivatives, and let $F_i$ denote the partial derivative of $F$ with respect to its $i$th argument.  This gives
\bea
-\frac{h_t}{J\cos\theta}=\int dx\int dy \Big\{&& F(-x,-y,\theta,0,0,0)+F_1(-x,-y,\theta,0,0,0)(yh_y\cot\theta)\nonumber\\  \nonumber
 && + F_2(-x,-y,\theta,0,0,0)(-xh_y\cot\theta)\\ \nonumber
 &&-\sec\theta\frac{\partial}{\partial \theta}\big[\cos\theta F(-x,-y,\theta,0,0,0)h_x\big]\\ \nonumber
 &&+K_{11}F_4(-x,-y,\theta,0,0,0)+K_{12}F_5(-x,-y,\theta,0,0,0)\\ 
 &&+K_{22}F_6(-x,-y,\theta,0,0,0)\Big\}.\label{full}
\eea

To simplify this expression, we will examine it term by term and employ Eq.~(\ref{hdef}).  The second term on the right-hand side of Eq.~(\ref{full}) is
\be
I_2\equiv \int dx\int dy F_1(-x,-y,\theta ,0,0,0)y\cot\theta (S_2+K_{12}x+K_{22}y).\label{above}
\ee
$I_2$ is in fact zero.  To see this, recall that we have assumed that the solid surface is amorphous. Independent of the details of the crater function $F(u,v,\phi,K_{11},K_{12},K_{22})$, therefore, symmetry demands that it be an even function of $v$ if $K_{12}=0$.  Thus, the terms which are proportional to odd powers of $y$ in the integrand of Eq.~(\ref{above}) integrate to zero.  The remaining term in the integrand vanishes upon integration over $x$ since
\be
\int dx F_1(-x,-y,\theta,0,0,0)=F(-x,-y,\theta,0,0,0)|_{x=-\infty}^{x=\infty}=0.
\ee

The third term on the right-hand side of Eq.~(\ref{full}) may be written
\be
I_3\equiv-\int dx\int dy F_2(-x,-y,\theta,0,0,0)x\cot\theta(S_2+K_{12}x+K_{22}y).
\ee
Again using the symmetry of $F$, we see that $F_2(-x,-y,\theta,0,0,0)$ is an odd function of $y$, and thus the terms in the integrand that are proportional to even powers of $y$ will integrate to zero.  This leaves
\bea
I_3&=&-\cot\theta K_{22}\int dx\int dy F_2(-x,-y,\theta,0,0,0) xy\nonumber\\
&=&\cot\theta K_{22}\int dx\int dy F(x,y,\theta,0,0,0)x\nonumber\\
&=&\cot\theta K_{22}M^{(1)}_x,
\eea
where we have integrated by parts and changed the dummy variables of integration from $x$ to $-x$ and from $y$ to $-y$.

The fourth term on the right-hand side of Eq.~(\ref{full}) is identical to the analogous term in the 2D case, except that $h_x$ now contains the additional term $K_{12}y$.  However, since $F(-x,-y,\theta,0,0,0)$ is an even function of $y$, this term makes no contribution.

Without additional assumptions or specific information about the crater function, the fifth and seventh terms on the right-hand side of Eq.~(\ref{full}) cannot be simplified further.  However, we may eliminate the dependence of $h_t$ on $K_{12}$ using a symmetry argument.  Notice that a surface described by $h(x,y)=K_{12}xy$ is invariant under the transformation $y\rightarrow-y$, $K_{12}\rightarrow-K_{12}$.  We may thus write
\be
F(x,y,\theta,0,K_{12},0)=F(x,-y,\theta,0,-K_{12},0).
\ee
It follows that $I_6$, the sixth term on the right-hand side of Eq.~(\ref{full}), is given by
\bea
\frac{I_6}{K_{12}}=\Bigg[\frac{\partial}{\partial K_{12}}&&\int _{-\infty}^{\infty} dx \int _{-\infty}^{\infty} dy F(-x,-y,\theta,0,K_{12},0)\Bigg]\Bigg|_{K_{12}=0}\nonumber\\
=\Bigg\{\frac{\partial}{\partial K_{12}}\Bigg[&&\int _{-\infty}^{\infty}dx\int _{0}^{\infty} dy F(-x,-y,\theta,0,K_{12},0)\nonumber\\
&&+\int _{-\infty}^{\infty}dx\int _{0}^{\infty} dy F(-x,-y,\theta,0,-K_{12},0)\Bigg]\Bigg\}\Bigg|_{K_{12}=0}.
\eea
The quantity in the square brackets in the later expression is an even function of $K_{12}$.  As a consequence, $I_6$ vanishes
and $C_{12}=0$.  We could have reached this conclusion \emph{a priori} from Eq.~(\ref{OGEOM}):  since the system is invariant under a reflection about the $x-z$ plane, $h_t$ must also remain invariant under this transformation, which implies that $C_{12}=0$. 

We define
\be
M_{K_{11}}=\int\int F(x,y,\theta,K_{11},0,0)dxdy\label{k11a},
\ee
\be
M_{K_{22}}=\int \int F(x,y,\theta,0,0,K_{22})dxdy\label{k22a},
\ee
and
\be
M_x^{(n)}=\int \int F(x,y,\theta,0,0,0)x^ndxdy.\label{m1xa}
\ee
Collecting terms, we arrive at a simpler form of Eq.~(\ref{full}),
\bea
-\frac{h_{t}(0,0,t)}{J\cos\theta}=\,&&M^{(0)}_{x}-S_1\sec\theta \frac{\partial}{\partial \theta} (\cos{\theta}M^{(0)}_{x})\nonumber\\&&+K_{11}\Bigg[\sec\theta\frac{\partial}{\partial \theta}\left(\cos\theta M^{(1)}_{x}\right)+\frac{\partial}{\partial K_{11}} M_{K_{11}}\Big|_{K_{11}=0}\Bigg]\nonumber \\
&&+K_{22}\Bigg[\cot\theta M_x^{(1)}+\frac{\partial}{\partial K_{22}}M_{K_{22}}\Big|_{K_{22}=0}\Bigg].\label{FULLEOM}
\eea
Comparing this with Eq.~(\ref{OGEOM}), we conclude that Eqs.~(\ref{C0}) - (\ref{C11}) remain valid, but the moments $M_{K_{11}}$ and $M_x^{(n)}$ are now given by Eqs.~(\ref{k11a}) and (\ref{m1xa}).  We also have found that $C_2=C_{12}=0$ and that
\be
C_{22}(\theta)=-\cos\theta\cot\theta M_x^{(1)}-\cos\theta\frac{\partial}{\partial K_{22}}M_{K_{22}}\Big|_{K_{22}=0}.\label{C22}
\ee
The first term on the right-hand side of Eq.~(\ref{C22}) is present because if $h_y$ is nonzero at the point of impact \textbf{P}, the local normal $\bm{\hat n}$
and the local downbeam direction $\bm{\hat t}_u$ have nonzero components along the $y$-direction.
The second term results from the explicit dependence of the crater function on the curvature in the $y$-direction. 

Despite the appearance of the factor of $\cot \theta$ in Eq.~(\ref{C22}), $C_{22}(\theta)$ is well behaved in the limit $\theta \rightarrow 0$.  To see this, note that for small $\theta$,
\be
M_x^{(1)}(\theta)\cong R_0+R_1 \theta,
\ee
where $R_0$ and $R_1$ are finite constants.  Symmetry demands that $M_x^{(1)}(0)=0$, and thus $R_0=0$.  Therefore, in the limit of small $\theta$, the lowest order term $M_x^{(1)}$ is proportional to $\theta$.  It follows that 
\be
\lim_{\theta \to 0} \left[\cos\theta \cot\theta 
M_x^{(1)}(\theta)\right]=R_1.
\ee
The value of the constant $R_1$ of course depends on the specifics of the crater being considered, but it is finite.

\section{Application of the Formalism to the Sigmund Model}
\label{sec:conclusions}

In this section, we demonstrate explicitly that our crater function formalism yields the exact BH coefficients for the Sigmund model.  
The crater function for the Sigmund model is given by Eq.~(8) of Ref.~[\onlinecite{Bradley11}].  For convenience, we will adopt the same notation that was used in that work.\cite{footnote2}  On average, an impact at the origin produces a crater whose negative depth at the point $\bm{r} = x\bm{\hat{x}}+y\bm{\hat{y}}+h(x,y)\bm{\hat{z}}$ is 
\bea
F(x,y,\theta,K_{11},K_{12},K_{22})=\frac{\epsilon \Lambda}{(2\pi)^{{3}/{2}}\alpha \beta^2}}{ \exp\Bigg(&&-\frac{1}{2 \alpha^2}[a-x\sin\theta+h(x,y)\cos\theta]^2\nonumber\\
&&-\frac{1}{2 \beta^2}[x\cos\theta+h(x,y)\sin\theta]^2-\frac{1}{2 \beta^2}y^2\Bigg).
\label{FSig}
\eea
If the distance between the origin and $\bm{r}$ does not exceed a few times $l$, then
we may set 
\be
h(x,y)= \frac{1}{2}K_{11}x^2+K_{12}xy+ \frac{1}{2}K_{22}y^2
\label{hcrat}
\ee
in Eq.~(\ref{FSig}).  The dependence of the crater for the Sigmund model on the components 
$K_{ij}$ of the curvature tensor becomes manifest once
Eq.~(\ref{hcrat}) has been inserted into Eq.~(\ref{FSig}).

For brevity, let
\be
D\equiv \frac{a^2\epsilon \Lambda}{(2\pi)^{{3}/{2}}\alpha \beta^2}.
\ee
We readily obtain
\bea
M^{(0)}_x&&=D\int \int \exp\Big(-\frac{1}{2}a_\alpha^2(1-x\sin\theta)^2-\frac{1}{2}a_\beta^2x^2\cos^2\theta-\frac{1}{2}a_\beta^2y^2\Big)dxdy \nonumber\\
&&=D e^{-a_\alpha^2/2}\int\int\exp\Big(-\frac{B_1}{2}x^2+Ax-\frac{a_\beta^2}{2}y^2\Big)dxdy\nonumber\\
&&=D e^{-a_\alpha^2/2}\frac{2\pi}{a_\beta \sqrt{B_1}}\exp\left(\frac{A^2}{2B_1}\right)
\label{M0}
\eea
and
\bea
M_x^{(1)}&&=aD e^{-a_\alpha^2/2} \int\int x \exp\left(-\frac{B_1}{2}x^2+Ax-\frac{a_\beta^2}{2}y^2\right)dxdy\nonumber\\
&&=\frac{aA}{B_1}M^{(0)}_x.
\label{M1}
\eea
To find $C_{11}$ and $C_{22}$, we need the partial derivatives of the curvature dependent moments $M_{K_{22}}$ and $M_{K_{22}}$ with respect
to $K_{11}$ and ${K_{22}}$, respectively.  Since the $K_{ij}$'s do not depend on $x$ and $y$, we may exchange differentiation with respect to the $K_{ij}$'s
with integration over $x$ and $y$.  This gives
\bea
\frac{\partial }{\partial K_{11}}&& M_{K_{11}}\Big|_{K_{11}=0}=\int\int \frac{\partial}{\partial K_{11}}F(x,y,\theta,K_{11},0,0)\Big|_{K_{11}=0}dxdy\nonumber\\
&&=-aD e^{-a_\alpha^2/2}\int \int\exp\left(-\frac{B_1}{2}x^2+Ax-\frac{a_\beta^2}{2}y^2\right)\left(\frac{B_{2}}{2} x^2+Cx^3\right)dxdy\nonumber\\
&&=-aM^{(0)}_x\left(\frac{A^2 B_2}{2B_1^2}+\frac{A^3 C}{B_1^3}+\frac{B_2}{2B_1}+\frac{3 A C}{B_1^2}\right).
\label{deriv1}
\eea
Similarly,
\bea
\frac{\partial }{\partial K_{22}}&& M_{K_{22}}\Big|_{K_{22}=0}=\int\int \frac{\partial}{\partial K_{22}}F(x,y,\theta,0,0,K_{22})\Big|_{K_{22}=0}dxdy\nonumber\\
&&=-aD e^{-a_\alpha^2/2}\int \int\exp\Big(-\frac{B_1}{2}x^2+Ax-\frac{a_\beta^2}{2}y^2\Big)\Big(\frac{B_{2}}{2} y^2+Cxy^2\Big)dxdy\nonumber\\
&&=-M^{(0)}_x\frac{a}{a_\beta^2}\Big(\frac{B_2}{2}+\frac{AC}{B_1}\Big).
\label{deriv2}
\eea
We must also compute the derivative of $ M^{(1)}_x \cos \theta$ with respect to $\theta$.  We obtain
\bea
\frac{\partial( M_x^{(1)}\cos\theta)}{\partial \theta}=&&\frac{\partial}{\partial \theta} \left(\frac{aA}{B_1}M^{(0)}_x\cos\theta\right)\nonumber\\
=&&-aM^{(0)}_x\Bigg[\frac{A\sin\theta}{B_1}
-\cos\theta\left(\frac{B_2}{B_1}+\frac{6AC}{B_1^2}+\frac{A^2B_2}{B_1^2}+\frac{2A^3C}{B_1^3}\right)\Bigg].
\label{deriv3}
\eea
Finally, we will need the identity
\be
\frac{1}{a_\beta^2}\Big(B_2+\frac{2AC}{B_1}\Big)=\cot\theta\frac{A}{B_1}.
\label{identity}
\ee
Inserting Eqs.~(\ref{deriv1}) and (\ref{deriv3}) into Eq.~(\ref{C11}) yields
\be
C_{11}(\theta)
=aM^{(0)}_x\Bigg[\frac{A\sin\theta}{B_1}
-\frac{\cos\theta}{2}\left(\frac{B_2}{B_1}+\frac{6AC}{B_1^2}+\frac{A^2B_2}{B_1^2}+\frac{2A^3C}{B_1^3}\right)\Bigg].\label{C11f}
\ee
Similarly, inserting Eqs.~(\ref{M1}) and (\ref{deriv2}) into Eq.~(\ref{C22}) and applying the identity (\ref{identity}), we have
\bea
C_{22}(\theta)
=&&-aM^{(0)}_x\Bigg[-\frac{1}{a_\beta^2}\Bigg(\frac{B_2}{2}+\frac{AC}{B_1}\Bigg)+\cot\theta\frac{A}{B_1}\Bigg]\cos\theta\nonumber\\
=&&-M^{(0)}_x\frac{a}{a_\beta^2}\Big(\frac{B_2}{2}+\frac{AC}{B_1}\Bigg)\cos\theta\label{C22f}.
\eea
Note as well that explicit expressions for $C_0$ and $C_1$ can be obtained by inserting Eq.~(\ref{M1})
into Eqs.~(\ref{C0}) and (\ref{C1}).  The resulting expressions for $C_0$ and $C_1$ and Eqs.~(\ref{C11f}) and (\ref{C22f}) for $C_{11}$ and $C_{22}$ agree with the results obtained by BH for the Sigmund model.

\section{Discussion}
\label{sec:discussion}

\subsection{Elemental Materials}

The key results of this paper are given by Eqs.~(\ref{C11}) and (\ref{C22}).  These equations give a means of computing the coefficients $C_{11}$ and $C_{22}$ if the curvature dependent crater function is known.
These coefficients play a key role in determining whether parallel-mode or perpendicular-mode ripples form
or if the surface remains flat.

In their 2011 paper, Norris {\sl et al.}\cite{Norris11}~gave explicit expressions for $C_{11}$ and $C_{22}$, namely
\be
C_{11}(\theta)=-\frac{d}{d \theta}\left[M_x^{(1)}(\theta)\cos\theta\right]\label{Norris1},
\ee
and
\be
C_{22}(\theta)=-M_x^{(1)}(\theta)\cos\theta\cot\theta.\label{Norris2}
\ee
Our results Eqs.~(\ref{C11}) and (\ref{C22}) show that Eqs.~(\ref{Norris1}) and (\ref{Norris2}) are
a good approximation only if the curvature dependence of the crater function is negligible.\cite{footnote} 

In the Sigmund model of ion sputtering, the form of the crater depends on the curvature of the surface at the point of impact, despite a statement to the contrary in Ref.~[\onlinecite{Norris09}].  This point
is discussed in detail in Ref.~[\onlinecite{Nietiadi13}].  The second terms on the right-hand sides of Eqs.~(\ref{C11}) and (\ref{C22}) therefore yield nonzero contributions to $C_{11}$ and $C_{22}$.  These contributions were computed explicitly in the preceding section.

For normal-incidence ion bombardment, the values of $C_{11}$ and $C_{22}$ obtained by neglecting the
curvature dependence of the crater function [Eqs.~(\ref{Norris1}) and (\ref{Norris2})] differ by a factor of two from the exact values for the Sigmund model [Eqs.~(\ref{C11f}) and (\ref{C22f})].
In fact, Norris {\sl et al.}'s result for $C_{22}$ is equal to twice the exact value for all angles of incidence
$\theta$.

To get an idea of how much Eq.~(\ref{Norris1}) 
differs from the exact result for the Sigmund model for nonzero values of $\theta$, see Fig.~2.  
The values of $a$, $\alpha$ and $\beta$ used in that figure are for 1~keV Ar$^+$ bombardment of silicon.\cite{Madi11}  The ratio of $C_{11}$ as given by Eq.~(\ref{Norris1}) to the exact value is greater than two for a broad range of $\theta$ values.  For $\theta=45^\circ$, for example, the ratio exceeds 3.5.  The angle where 
the switch from parallel- to perpendicular-mode ripples occurs is $50.8^\circ$ but, if we use Eqs.~(\ref{Norris1}) and (\ref{Norris2}), this angle is found to be $66.7^\circ$, fully $15.9^\circ$
higher than the correct value. 

\centerline{\includegraphics[width=4.0in]{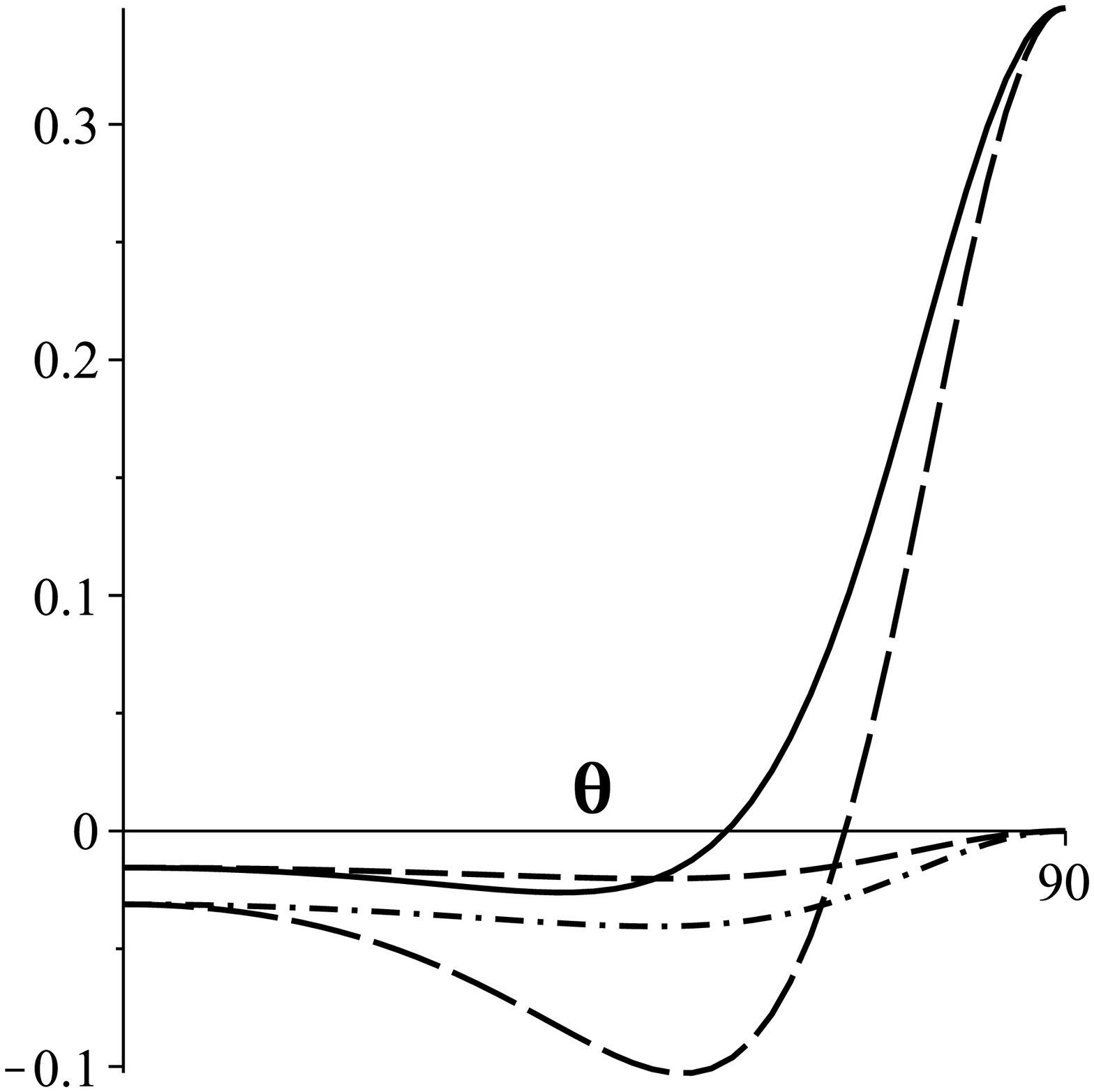}}

\begin{figure}[h]
\caption{
The coefficients $C_{11}$ and $C_{22}$ as functions of $\theta$ for the Sigmund model.  The exact results for $C_{11}$ and $C_{22}$ are shown with a solid and dashed curve, respectively.  The results obtained for $C_{11}$ and $C_{22}$ if the curvature dependence of the crater function is neglected are shown with long dashes and with a dash-dotted curve, respectively.  The values of $a$, $\alpha$ and $\beta$ employed are 
for 1 keV Ar$^+$ bombardment of silicon.  The coefficients are in units of 
$2\sqrt{2\pi}/(\Lambda\epsilon a_\alpha^3 a_\beta)$ and $\theta$ is given in degrees.
}
\label{figure2}
\end{figure} 

Recently, Nietiadi and Urbassek carried out MD simulations of 500 eV Ar$^+$ bombardment of an amorphous silicon target.\cite{Nietiadi13}  They found that the craters for curved surfaces 
are substantially different than those for a flat surface.  These observations and our results for the
Sigmund model lead us to the conclusion that the errors incurred by neglecting the curvature
dependence of the crater function are typically not small.

Perkinson \emph{et al.}\cite{PerkinsonXX}~have pointed out some apparent inconsistencies in the coefficients given by Norris \emph{et al.}, Eqs.~(\ref{Norris1}) and (\ref{Norris2}).  Because Norris \emph{et al.}'s expression for $C_{11}$ is a derivative with respect to $\theta$ of a function which vanishes at $\theta=0$ and ${\pi}/{2}$, the integral of $C_{11}$ is
\be
\int_{0}^{{\pi}/{2}}  C_{11}(\theta)d\theta=- M_x^{(1)}(\theta)\cos\theta\Big|_{\theta=0}^{\theta={\pi}/{2}}=0\label{anz1}
\ee
for their theory.  Additionally, using Norris \emph{et al.}'s expressions (\ref{Norris1}) and (\ref{Norris2}), we obtain
\be
C_{11}(\theta)=\frac{d}{d \theta}\big(C_{22}\tan\theta\big).\label{anz2}
\ee
Perkinson \emph{et al.}'s experiments provide evidence that the actual values for $C_{11}$ and $C_{22}$ do not satisfy either Eq.~(\ref{anz1}) or (\ref{anz2}).  This again suggests that the errors incurred by neglecting the curvature dependence of the crater function are significant.  When one includes the curvature dependence, however, Eqs.~(\ref{C11}) and (\ref{C22}) result, and
Eqs.~(\ref{anz1}) and (\ref{anz2}) do not apply.  Our 
CFF therefore does not suffer from the same difficulties as that of Norris \emph{et al.}

If the correct formulae (\ref{C11}) and (\ref{C22}) are to be used in combination with MD simulations to obtain accurate estimates 
of $C_{11}$ and $C_{22}$, it is not sufficient to find the crater function for a flat surface.  Instead, to find $C_{11}$, craters on a curved surface of the form
$h(x,y)=K_{11}x^2/2$ must be found for a range of small values of $K_{11}$ so that the derivative $\partial M_{K_{11}}/\partial K_{11}$ can be
computed for $K_{11}=0$.    Naturally, an analogous statement applies to determining $C_{22}$.  In that case, craters on a surface that has the form $h(x,y)=K_{22}y^2/2$
are needed.  This means that the computational resources necessary to find accurate values of the coefficients
$C_{11}$ and $C_{22}$ are considerably greater than it was previously thought.

As Eq.~(\ref{C0}) shows, the crater function for a flat surface is all that is needed to compute $C_0$.  Our expression for $C_0$ agrees with that of Norris \emph{et al.}~as a consequence.  Our extended CFF also yields an expression for $C_1$, Eq.~(\ref{C1}).  Norris \emph{et al.} did not give an explicit formula that relates $C_1$ to a crater function moment.\cite{Norris11}  This coefficient is needed if one wishes to find the velocity with which parallel-mode ripples propagate over the solid surface.

Norris, Brenner and Aziz introduced their CFF in 2009 but did not apply it to estimate 
the coefficients in the EOM for a particular target material or choice of ion beam.\cite{Norris09}  Subsequently,
Norris {\it et al.}~carried out MD simulations of the bombardment of a flat silicon surface with 100 and 250 eV Ar$^+$ ions and then used Eqs.~(\ref{Norris1}) and (\ref{Norris2}) to obtain
estimates of $C_{11}$ and $C_{22}$.\cite{Norris11} They also divided the first moment of the crater function $M_x^{(1)}$ into erosive and redistributive parts and so determined the relative contributions of the curvature dependence of the sputter yield and mass redistribution to $C_{11}$ and $C_{22}$.  This led to their conclusion that mass redistribution is predominant.

As we have seen, Eqs.~(\ref{Norris1}) and (\ref{Norris2}) lead to substantial errors in the
values of $C_{11}$ and $C_{22}$ for the Sigmund model.  For normal incidence, the common value of
$C_{11}$ and $C_{22}$ obtained by neglecting the curvature dependence of the crater function is twice as large as the correct value, for example.  Although Eqs.~(\ref{Norris1}) and (\ref{Norris2}) lead to an overestimate
of the erosive contribution to $C_{11}$ and $C_{22}$ in the case of the Sigmund model, it is likely
that for some other types of craters these formulae would result in a substantial underestimate of the
erosive contribution.  Norris {\sl et al.}'s assertion that mass redistribution has a substantially greater effect
than curvature dependent sputtering must therefore be treated with caution.

Although the results of Norris {\sl et al.}~were confined to bombardment of silicon with a low energy argon ion beam, they claimed that mass redistribution is predominant and that the curvature dependence of the sputter yield is \lq\lq essentially irrelevant'' for all ion species, angles of incidence and energy, and for all choices of target material.  In our view, this is overreaching.  Silicon and other semiconductors are readily amorphorized by ion bombardment, making ion-induced flow relatively easy.  Mass redistribution might therefore be much more important for these materials than it is for others.  Moreover, below a threshold energy, no sputtering occurs.  Accordingly, by confining their attention to low-energy ions, Norris {\sl et al.}~chose to study a regime in which erosive effects are more likely to be dominated by mass redistribution.
The relative importance of curvature dependent sputtering and mass redistribution may be reversed
as the ion energy is increased.  Indeed, Hofs\"ass has carried out binary
collision Monte Carlo simulations and has found that the 
curvature-dependence of the sputter yield is the dominant contribution
to the pattern formation, except for very low energy irradiation of a light target material with heavy ions.\cite{Hofsass13}

\subsection{Binary Materials}

In 1999, a series of fascinating experiments by Facsko {\sl et al.}~revealed that normal-incidence bombardment of the binary compound GaSb with an argon ion beam can produce a densely packed, 
highly regular hexagonal array of nanodots.\cite{Facsko99}  Bradley and Shipman (BS) subsequently introduced a theory that accounts for the formation of orderly hexagonal arrays of nanodots
when the flat surface of a binary compound is subjected to normal-incidence ion bombardment.\cite{Bradley10,Shipman11,Bradley12}
In their theory, the coupling between the topography of the surface and a thin surface layer of altered composition is
the key to the observed pattern formation.  In addition, in the BS theory, the curvature dependence of the sputter yields is responsible for the instability 
that leads to the formation of the nanodots.

In an effort to test the BS theory, Norris and co-workers extended their CFF to binary materials.\cite{NorrisXX}  They then performed MD simulations to find the craters
produced by argon ion bombardment of GaSb and input the results into their CFF.  The resulting estimated parameter values do not lead to the formation of 
ordered arrays of nanodots in the BS model.  These results have led Norris to suggest that the instability that leads to the formation of the nanodot arrays stems from phase segregation
rather than the curvature dependence of the sputter yields, and to generalize the BS theory to include the former effect.\cite{Norris13}

Norris {\sl et al.}~neglected the curvature dependence of the crater function in extending their CFF to binary target materials.\cite{NorrisXX}  Accordingly, it is possible  that the errors in their estimated parameter values are quite large, and that improved estimates would in fact lead to the emergence of ordered arrays of nanodots from the BS model.  Additional analytical work is needed in which the curvature dependence of the crater function
is taken into account when the CFF is extended to binary materials.  MD simulations that yield the crater function for curved GaSb surfaces would then permit significant improvements in the estimated parameter values, and would indicate whether curvature dependent erosion or phase segregation is responsible for the formation of the nanodots.

\section{Conclusions}
\label{sec:conclusions}

In principle, the crater function $F$ depends on the entire shape of the surface in the vicinity of the point of impact.
In this paper, we extended the crater function formalism to include the dependence of $F$  
on the curvature of the surface at the point of impact.  Explicit expressions for the constant coefficients in the continuum equation of motion were derived; these reduce to the results given by Norris {\it et al.}\cite{Norris11}~\emph{only} if the curvature dependence of
the crater function is negligible.  Our extended crater function formalism yields the exact coefficients 
for the Sigmund model of ion sputtering.   In contrast, if the curvature dependence of the crater function
is neglected, substantial errors in the estimated values of the coefficients typically ensue.  

Our results show that accurately estimating the coefficients in the equation of motion using craters obtained 
from molecular dynamics simulations will require significantly more computational power than was previously thought.  They also lead us to question the reliability of 
the coefficient estimates that lead to the
recent claim that sputtering is relatively unimportant for ion-induced pattern formation and that mass redistribution is always predominant.  

In future work, we will include the curvature dependence of the crater function in extending the crater function formalism to binary materials.  Once this has been done, it will become possible to make reliable estimates of the coefficients in the equations of motion.  This will likely lead to greater insight into the physical origin of the highly ordered hexagonal arrays of nanodots that sometimes develop during normal-incidence ion bombardment of a binary material.

\begin{acknowledgments}

We have benefited from stimulating discussions and correspondence with Hans Hofs\"ass, Karl F. Ludwig, Scott A. Norris and Patrick D. Shipman.  R.M.B.~would like to thank the National Science Foundation for its support through grant DMR-1305449.

\end{acknowledgments}

\vfill\eject


\begin{thebibliography}{00}
\bibitem{footnote2} Note that the $\theta$ used in Ref.~[\onlinecite{Bradley11}] is minus the $\theta$ employed here and in Ref.~[\onlinecite{Norris11}].
\end{thebibliography}
\end{document}